\documentclass[conference]{IEEEtran}
\IEEEoverridecommandlockouts
\usepackage{cite}
\usepackage{amsmath,amssymb,amsfonts}
\usepackage{algorithmic}
\usepackage{graphicx}
\usepackage{textcomp}
\usepackage{xcolor}
\usepackage{paralist}
\usepackage[linesnumbered,ruled,vlined]{algorithm2e}

\SetCommentSty{mycommfont}
\def\BibTeX{{\rm B\kern-.05em{\sc i\kern-.025em b}\kern-.08em
		T\kern-.1667em\lower.7ex\hbox{E}\kern-.125emX}}
\begin{document}
	
	\title{Incentivising Building Data Availability and Accessibility  Using Tokenized Data Assets\\}

	\author{\IEEEauthorblockN{ Sarad Venugopalan and Heiko Aydt}
		\IEEEauthorblockA{\textit{Singapore-ETH Centre} \\
			sarad.venugopalan@sec.ethz.ch, heiko.aydt@sec.ethz.ch}
	}
	
	\maketitle
	
	\begin{abstract}
		
		Smart cities are data driven and collect data from a variety of sources.  
		Certain  types of data such as building data  is  under-represented and remains harder to find despite its  value.
		Our goal is to incentivise the stakeholders to make building data easier to avail by turning it into an asset.
		We  use  tokenized building data assets on a blockchain to improve data accessibility. This is achieved by connecting building data owners with the consumers of building information via tokens (fungible and non-fungible), which serves the purpose of coordinating the activities of the built ecosystem.
		Further, we  present our system architecture designed to sustain the economic incentives for interested parties and individuals.
		
	\end{abstract}

	
	\section{Introduction}
	
	In construction, a substantial amount of energy is spent in
	extracting the materials out of the earth,  processing  and transporting
	to its destination.  It  is not limited to the energy expended on the material  at a construction site but also includes all the energy spent preceding it.
	Its grey energy~\cite{Hill2021} is viewed as the  `hidden energy' stored in the  end product.
	When reusable and recyclable material from demolished buildings are not used for its intended purpose,
	it adds strain on  depleted resources (as a result of extensive material mining~\cite{Swilling2018}), to be found anew. Scarce
	resources becomes expensive, the energy required to process it may increase (in periods of increased energy costs), and has cumulative environmental
	effects that add costs to the health and lives of individuals and societies.
	Due to the scale of urban construction and the material that needs to be found, processed and shipped to the destination, the construction materials have a huge impact on sustainability~\cite{UNEP2019} due to material scarcity, energy expended and environmental effects.
	Presently, the means to find reusable and recyclable material, its quality, quantity, deconstruction information, location and availability data are  inadequate.
	While  reuse and recycle has improved in some countries, a considerable part of it  ends up in landfills~\cite{USEPA2023,USEPA2020}.
	According to the US Environmental Protection Agency~\cite{USEPA2022}, “Materials and waste exchanges are markets for buying and selling reusable and recyclable commodities. Some are physical warehouses that advertise available commodities through printed catalogs, while others are simply websites that connect buyers and sellers. Some are coordinated by state and local governments; others are wholly private, for-profit businesses".
	This process is semi-automated and  operates in silos, making it difficult to find and optimise  resource allocation across the spectrum.
	
\begin{figure}
		\centering
		\includegraphics[width=0.9\linewidth]{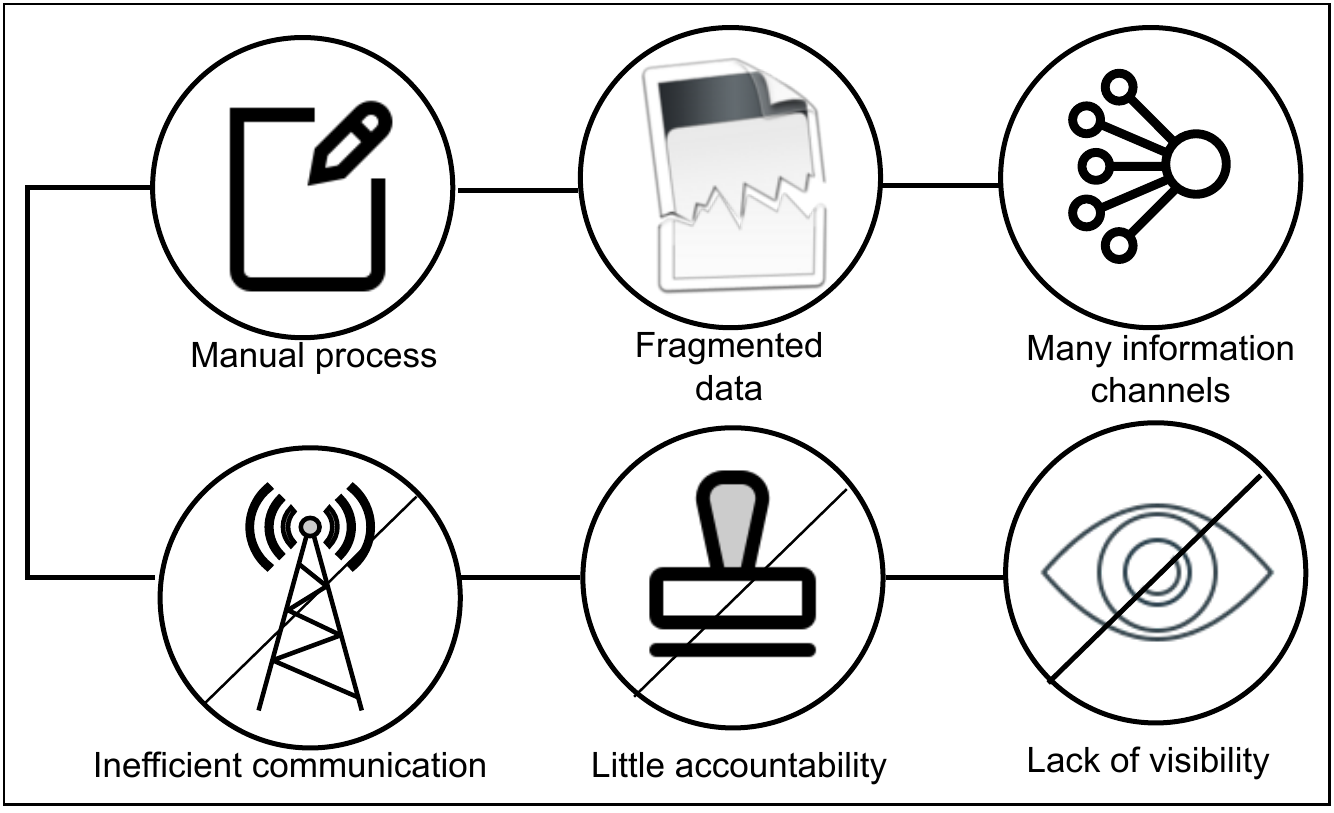}
		\caption{The difficulty in collecting construction data~\cite{Holtmann2019} arises from multiple channel data inputs from numerous people without redundancy and coherence checks, collection of fragmented data, and using a manual process to record it, coupled with inefficient communication, little accountability and a lack of data visibility. The use of different disconnected software widens the problem of working in silos and prevents real-time data availability.   }
		\label{fig:datacollection}
	\end{figure}

	Secondly, the non-/incomplete availability of building data to create building models and data analytics for research remains an issue.
	Building models at an urban scale are used to understand the past and current demand patterns, to test
	future scenarios and assist decision makers with making the best choices for existing and future cities.
	While building information is an essential input to create building models, it has missing data and rarely has the granularity required~\cite{Pei2022,Seif2023}. Some of the present methods to collect  the required data include obtaining it  from available datasets. This may be  open data from  government sources, such as municipalities, policy makers  and utilities. Others are
	indirectly generated data based on available data, such as,  using machine learning~\cite{YANG2021} to fill the missing gaps.
	There is a growing recognition that public policy
	needs to be  tested and evaluated based on evidence~\cite{Kohler2009}.
	According to Kohler and Hassler~\cite{Kohler2002},
	“The need for more reliable models of the
	composition and dynamic of the building stock
	has been recognised in most countries, however
	there is insufficient statistical data upon which
	to base them. As a result, research does not
	advance as fast as it should."
	In our work, we add to the suite of data collection techniques in the construction industry (see Section~\ref{sec:related}) by focusing on collecting data at the source and using incentivisation to build a sustainable economic model. Our built ecosystem is expected to ease some of the difficulties surrounding the collection of construction data (see Fig.~\ref{fig:datacollection}), and improve the ability to find reusable and recyclable building material, and  wholesome data  at the required granularity for researchers and analysts.
	
	We make the following contributions,
	\begin{compactenum}

		\item [\textit{i}.)] We present our architecture (see Section~\ref{sec:sysarch}) used to incentivise the stakeholders of the built ecosystem for data availability and accessibility.
		
		\item [\textit{ii}.)] We discuss the sustainability of our incentive model (Section~\ref{ssec:incentives}), provide application use cases (Section~\ref{ssec:usecase}) and justify the use of a blockchain (Section~\ref{ssec:blockchainjustification}) in our ecosystem.
	\end{compactenum}
	
	The rest of the paper is organised as follows.
	Section~\ref{sec:prelims} provides the preliminaries.
	Section~\ref{sec:Overview} presents the solution overview.
	Section~\ref{secsanalysis} details the system model and design goals.
	The system architecture is presented in Section~\ref{sec:sysarch}.
	Section ~\ref{sec:discussion} discusses the implications of our architecture.
	The related work is presented in Section~\ref{sec:related}.
	We conclude with our findings and contributions in Section~\ref{sec:conclusions}.

	\section{Preliminaries}
	\label{sec:prelims}
	
	In this section we provide definitions and requirements for data assets, building data assets and tokens.
	\subsection{Data Assets}
	\label{ssec: DataAssets}
	If an asset is “a resource with economic value that an individual, corporation, or country owns or controls with the expectation that it will provide a future benefit"~\cite{Barone2022}, then a data asset is a (digital and intangible) resource that an individual, corporation, or country owns or controls with the expectation that it will provide a future benefit. In order for data to be considered an asset it needs to satisfy the following requirements.
	\begin{compactenum}
		
		\item [(R1)] It must be impossible to arbitrarily duplicate a data asset - this turns data into a (unique) and identifiable object. It also introduces the element of scarcity.
		
		\item [(R2)] It must be possible for the owner(s) of a data asset to take possession and assume exclusive control (with possible exception of a regulatory authority) - this allows assets to become private property (subject to the laws of the jurisdiction the asset may be subject to) enabling its owners to accumulate, hold, delegate, rent, or sell their property.
		
		\item [(R3)] It must be possible for a data asset to be subject to a regulatory authority (if required) - this gives the data object legitimacy in the context of any relevant jurisdiction in the physical world.
		
		\item [(R4)] It must be possible to audit a data asset (e.g., in order to verify its integrity and test for any impairments).
		
		\item [(R5)] It must be possible to generate/derive and extract useful information from a digital asset (or parts thereof) that can be purchased/licensed and consumed by third parties to their benefit, thus generating a source of revenue.
		
	\end{compactenum}
	
	\subsection{Building Data Assets}
	\label{ssec: BDAdef}
	A Building Data Asset (BDA) is a data asset plus a domain-specific Application Program Interface (API) to interact with and manipulate the data in ways meaningful to the domain of interest (i.e., the building domain in case of BDAs).
	There can be many APIs within a domain. For example, within the building domain, one may have different APIs for different use cases such as power consumption (monthly power usage in an apartment unit), construction (recyclable material used~\cite{Honic2021}), and others. The same data asset may be accessed/interacted  through different APIs, whereby each API only interacts with some parts of the underlying data asset.
	
	\subsection{Tokens}
	\label{ssec:tokens}
	
	A token is a standardised unit of accounts. A token represents a particular interest in some aspect of the data asset and is used to coordinate the activities of the business. For example, it may represent a share of ownership in the data asset, or right to use.
	Typically, a token represents a data asset created on a  blockchain to enable its properties. When stored on a blockchain, the  asset does not contain the bulk of the associated data stored along with it, instead only a link to it. When tokens are issued on the blockchain, and correspond to the properties of the data asset, they are called tokenized assets. It may be fungible or non-fungible, depending on its application.
	
	\begin{figure*}
		\centering
		\includegraphics[width=1.0\linewidth]{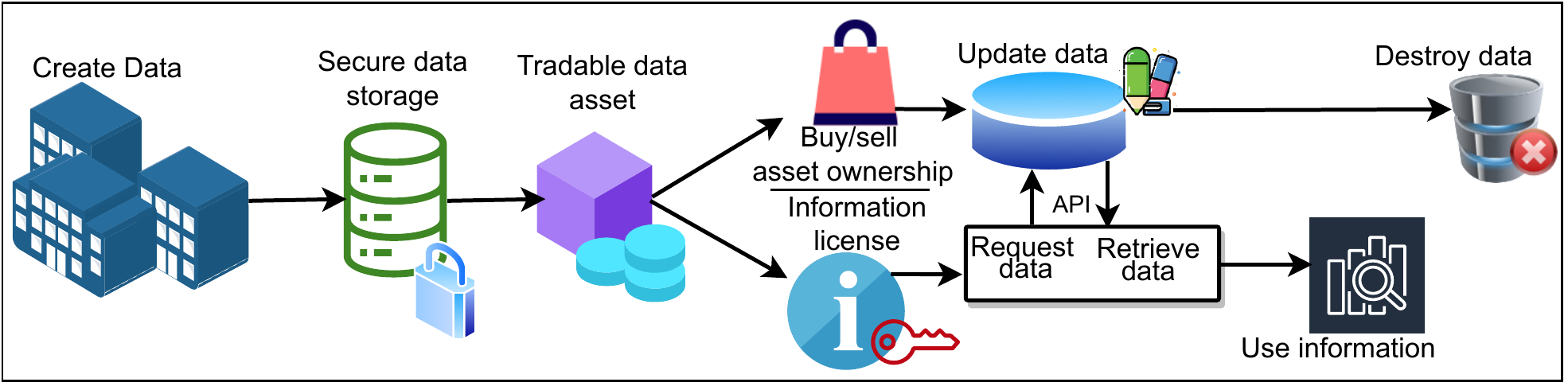}
		\caption{ Building data is first created and securely stored before it is turned into a tradable asset. Our economic model consists of i) trading the ownership of data and ii) charging the consumer a licence fee for the use of information. The owner may sell the information for profit and the consumer may use it for analysis and actionable insights. The data in storage may be updated over time by the owner, and reused by consumers. In general, the data is destroyed when the building is demolished, but may be retained if it has  historic value.  }
		\label{fig:lifecycle}
	\end{figure*}

	\section{Solution Overview}
	\label{sec:Overview}

	Building information  is currently not treated as something valuable beyond its immediate scope of application. One of the reasons for this is the lack of a long term value proposition and/or clear pathways for generating revenue using building data.
	In order to monetise building data, it has to be first systematically generated or collected (e.g., during the construction phase of a building). Furthermore, building data must be maintained, curated and regularly updated to remain relevant and up-to-date as time passes. All this incurs costs and companies may not see the point of doing so beyond what is needed to achieve their immediate tasks at hand.
	For example, design, construction and engineering companies may create and use building data (e.g., design and component databases, de-/construction details~\cite{Sanchez2021}) during the various phases of building construction. However, they may have no further use for this data after the building is constructed. However, ideally this data would be made available to businesses that specialise in operating and maintaining a building, who could also maintain and enhance the building data (e.g., with data from operations - such as energy consumption). Depending on the nature of specialisation, such businesses may also be able to monetise the data by selling insights (e.g., about maintenance needs/events) to interested third parties (e.g., building operators/owners, maintenance contractors, tenants, research and analytics companies). Yet other businesses may specialise in the mining of urban materials or by providing the necessary insights (e.g., what materials can be found where) to interested third parties - thus contributing to a circular economy~\cite{Cetin2021} of construction materials.
	
	We consider  the possibility if the companies (e.g., construction, design and engineering companies) that generate/collect data as by-product and that do not have the necessary infrastructure, skill (or interest for that matter) to monetise this data beyond their own use of the data
	could sell/transfer their building data to interested third parties in a standardised and structured manner - just like any other asset. This may allow businesses to emerge that specialise in acquiring (and aggregating) certain building data to generate revenue by exploiting valuable insights or generating revenue by selling them. While technical standards for building data itself may already exist (e.g., Building Information Modelling), there is generally a lack of standards and methods for addressing the economic/financial aspects of data.
	
	Pathways for generating revenue (see Fig.~\ref{fig:lifecycle}) would provide the necessary incentives to maintain and update data until the building is demolished (and possibly beyond). Building data assets have a crucial role to play in the context of sustainable material flows in circular future cities~\cite{Heiko2022,Hunhevicz2022}. Among other potential use-cases for building data assets, they may provide information about building components and their materials to entities that specialise in mining and recycling building materials, thus contributing to a circular building materials economy.
	
	\section{System Model and Design Goals}\label{secsanalysis}
	
	\subsection{System Model}
	\label{ssec:sysmodel}
	We identify the following stakeholders and service providers (see Fig.~\ref{fig:ecosystem}).
	\begin{compactenum}
		
		\item \textbf{Contractors} are people tasked to collect the required data (e.g., building data or electricity data), digitise it in a prescribed format, and supply it to the building owner.
		
		\item \textbf{Data asset owner(s)} are  people who have ownership of the building data.  The data asset owners are building owners or those with close relationships with the building. Asset owners may employ contractors one-off, or periodic. Typically, data assets are owned by the building owner or bought by an information asset management company from a building owner.
		
		\item \textbf{Data certifier(s)} are one or more competent and authorised person(s) who physically verifies the correctness of building data (or its components) on behalf of the asset owners, who then  digitally signs the timestamped data (not shown in Fig.~\ref{fig:ecosystem}).
		
		\item A \textbf{tokenizer  company/entity} acts as a bridge between the physical and blockchain world. They are responsible for tokenizing data assets and providing those tokens to the building owner.
		
		\item \textbf{Data storage providers} are used to service the storage requirements. Two types of data storage are used --- cloud storage and public blockchain.
		Cloud storage services are hosted/rented on a subscription basis. It holds the datastores.
		The data asset is tokenized on a public blockchain.
		
		\item \textbf{Information users/consumers}  may be any individual, group, or organisation who use the data held by data asset owners via queries to profitably analyse it for useful insights. The data ownership is retained with the asset owner. The consumers are licensed to use the information but not sell it. The consumers pay for the information received.
		
		\item \textbf{Investors} may be any member of the public (individual or institutional) who buys tokens for dividends (royalties) and trades them.
		
	\end{compactenum}

	\subsection{Design Goals}
	The following are the main design goals.
	
	\subsubsection{Unhindered Access and Transfer of Tokenized Assets}
	BDAs should not be locked into a ‘permissioned’ platform to facilitate unhindered transfer between stakeholders, not limited to stakeholders that are part of any particular group, consortium or otherwise.
	This is to avoid operating in silos.
	
	\subsubsection{Censorship and Tamper resistance of Tokenized Assets}
	BDAs should be highly censorship resistant, i.e., it should be resistant to denial  attacks (such as denial of service~\cite{Msisac2022} or transaction censorship). Tamper resistance is to thwart unauthorised modifications.
	
	\subsubsection{Atomicity of Tokenized Assets}
	An asset owner may hold multiple assets related to materials and services in a building. Each asset is atomic.  For example, two  different special interest groups combined into one might not find the combined information they need under a single asset or data component.  They may have to query each of the two assets separately. This partitioning simplifies the  management of asset economics.
	
	\begin{figure}
		\centering
		\includegraphics[width=1.0\linewidth]{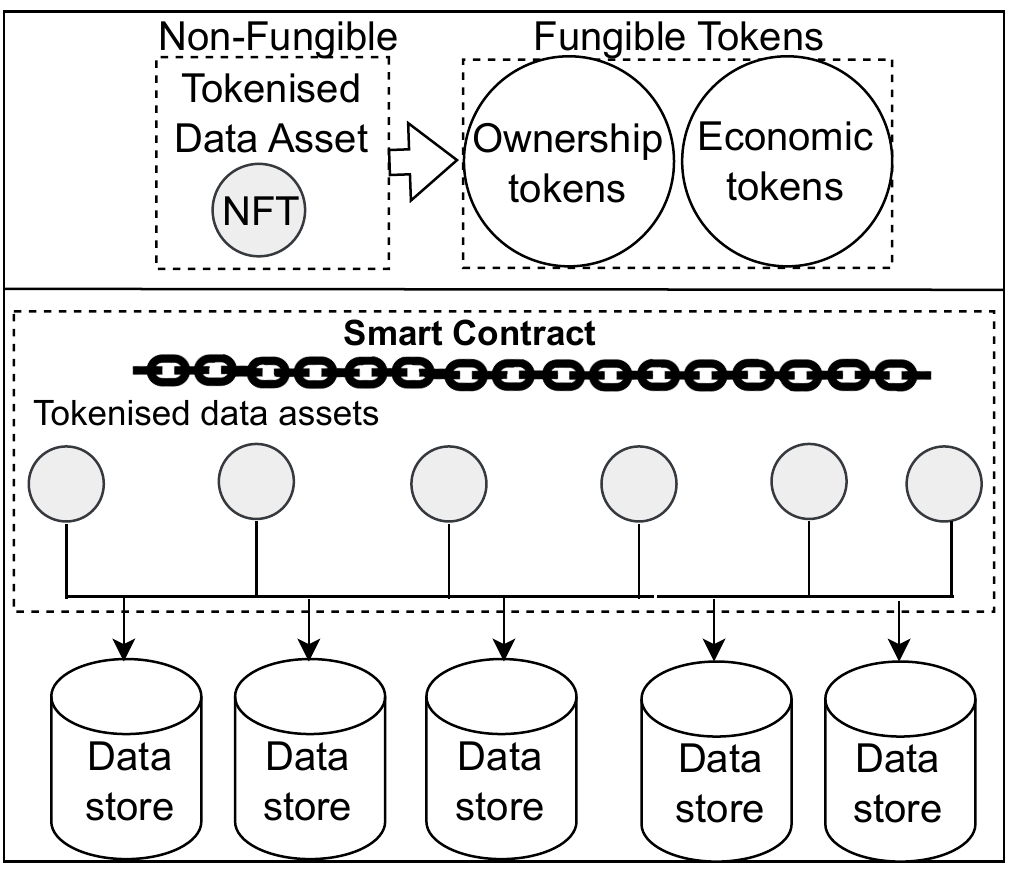}
		\caption{Each data asset is tokenized into an NFT. An NFT uniquely represents the data asset. Corresponding to each NFT, there are 2 types of fungible tokens, namely, ownership and economic tokens.  Every NFT points to its storage location in a datastore.}
		\label{fig:shares3}
	\end{figure}
	
	\subsubsection{Security of Data Assets}
	Providing security is necessary for two reasons. First, a data asset is of value only when its record keeping can be secured from ownership tampering, illegal alteration and access to trade and indisputable settlement of the asset. This is essential in a public setting where information can be bought/sold. The second reason is that only data that is owned and controlled can be sold.
	
	\subsubsection{ Economic incentives to keep the wheels moving} For services to be provided and maintained, we primarily rely on sustainable incentive mechanisms for cooperation and tokens for coordination, instead of  altruism or regulatory enforcement (which might result in the very minimum done to stay compliant).
	
	\section{System Architecture}
	\label{sec:sysarch}
	
\begin{figure*}
		\centering
		\includegraphics[width=1.0\linewidth]{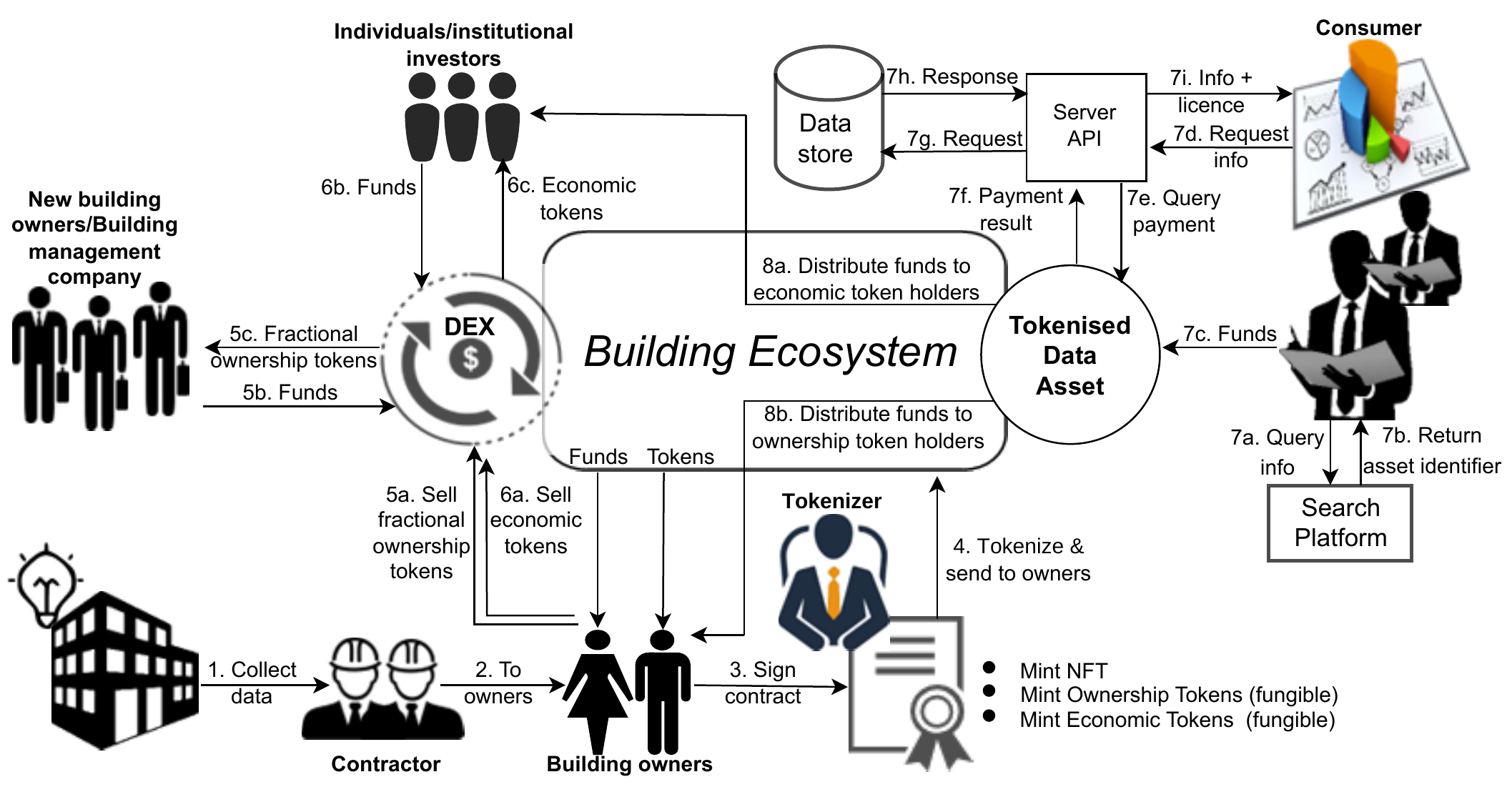}
		\caption{ The sequence of steps in bootstrapping and operating  the building ecosystem --- 1. Contractors collect building data and 2. is passed to the  building owner. 3. Building owners sign a legally binding  contract with the tokenizer. 4. Tokeniser issues/mints all the tokens agreed in the legal contract  and sends it to the owners. Owners use a decentralised exchange (DEX) to sell tokens, 5. either to sell a part of the ownership tokens, or 6. to raise capital using economic tokens. Further, the consumers interested  in the building data are able to 7a. query on a search platform for the 7b. asset  identifier (NFT contract address) of the tokenized data asset pointing to the required information. Once determined, 7c. the consumer pays the correct NFT contract address on the blockchain. 7d. Next, the consumer requests the server  for the information in the datastore. 7e. The server queries the NFT contract  to verify if the required payment was made. 7f. On confirmation of payment,
			7g. the server requests for the data identified by the NFT contract address from the datastore, and 7h. the store responds with the data. Based on the query,	7i. the requested information and licence are released to the consumer. The payment made by the consumer to  tokenized data asset (NFT) is step 7c., is proportionally distributed  between 8a. economic and 8b. ownership token holders, based on the rules in the NFT smart contract.
		}
		\label{fig:ecosystem}
	\end{figure*}

	In this section, we present the components and its interconnections for our building data asset ecosystem. We also introduce the tokens used for coordination between its stakeholders.
	Fig.~\ref{fig:shares3} shows the tokens used and Fig.~\ref{fig:ecosystem} presents the architecture.
	The basis for cooperation are incentives, as  for coordination are the tokens~\cite{Lamberty2020,Hunhevicz2022}.
	We demonstrate the interplay between incentives and tokens  (i.e.,  cooperation and coordination) using building assets.

	\subsection{Ecosystem Tokens}
	In our building ecosystem, tokens are used to bridge the gap between the buyer (data consumer) and seller (building data owner). It allows scarce but valuable building data made available (via mutual cooperation) by incentivising the owners.
	Our building ecosystem uses 3 types of tokens --- one is non-fungible and the remaining two are fungible (see Fig.~\ref{fig:shares3}).
	\begin{compactenum}
		\item \textit{Non-fungible token.} A NFT uniquely identifies the data asset. The two fungible tokens corresponding to the NFT are ownership and economic tokens.
		
		\item \textit{Ownership tokens.} They are fungible and it is used to facilitate fractional ownership sales. The sale of ownership (or part of it) may arise when a building or its units change owners and/or when a new building data asset management company buys information rights to the building.  
		Any data asset may have one or more owners, and the rule of agreement by a combined majority  of ownership token holders determines the rights (dispute resolution) for the data asset.
		Ownership tokens are sold only to other building owners and those who are closely associated with the building (eg., building data asset management company).
		
		\item \textit{Economic tokens.} They are fungible and may be sold by building data owners to individual/institutional investors on a public blockchain to raise capital or to cover the costs of tokenizing the asset and maintaining it (including paying the contractors and for storage). The economic token holders receive royalties for holding this token, and may also be traded on a decentralised exchange for profit.
		
	\end{compactenum}
	
	\subsection{Ecosystem bootstrapping and interconnections}
	The sequence of steps required to bootstrap the building ecosystem and operate it are shown in Fig.~\ref{fig:ecosystem}. First, the building owners hire contractors to collect building data (step 1, Fig.~\ref{fig:ecosystem}), who then pass it to owners in the prescribed digital file format (step 2, Fig.~\ref{fig:ecosystem}). The data is certified and digitally signed by an authorised expert on behalf of the owners (not shown in Fig.~\ref{fig:ecosystem}).  Further, the building owners are not expected to be tech-savvy. They use the services of a tokenizer  company/entity  to issue their tokens.
	The owner signs a legally binding  contract (step 3, Fig.~\ref{fig:ecosystem}) with the tokenizer. Based on this contract agreement, the tokenizer mints\footnote{Any tokenizer entity may be used. However, the ecosystem only accepts onboarding of data assets that follow the same set of rules, agreed  by its stakeholders. The contract with the tokenizer is legally binding to prevent them from arbitrarily minting tokens. Unlike cryptocurrencies (which have no jurisdiction), a building data asset is always intrinsically connected with a physical building and that building resides in a jurisdiction. Hence, it makes sense for the data assets for this physical asset to be under the same regulatory authority.} (issues) the tokens agreed in the signed contract, and sends it to the owners (step 4, Fig.~\ref{fig:ecosystem}). The ownership tokens\footnote{Further restrictions may apply to the sale of ownership tokens, as they are only intended to be sold to other building owners, or those closely associated with the physical building.
		This may be a condition of the legal contract between the building owner and tokenizer. Since the buildings are located in a physical jurisdiction, so are  the laws that govern it.
		The DEX only acts as a platform for trustless exchange of tokens.} and economic tokens may be sold on a decentralised exchange (DEX) (step 5 \& 6, Fig.~\ref{fig:ecosystem}), attached to the building ecosystem, for funds.
	Further, the information on the datastore pointed to by the data asset (NFT) needs to be searchable without revealing bulk of its information. This is achieved through an online custom search platform that is part of the ecosystem. It allows any interested consumer to query (step 7a, Fig.~\ref{fig:ecosystem}), for which it returns tokenized data asset (NFT) identifier (step 7b, Fig.~\ref{fig:ecosystem}).
	After the asset identifier is retrieved, the consumer is able to pay the correct NFT contract on the blockchain (step 7c, Fig.~\ref{fig:ecosystem}). Once payment is confirmed
	and the user is authenticated,  the request is granted to retrieve the required information from the datastore (step 7d-7i, Fig.~\ref{fig:ecosystem}).
	For the funds paid by the consumer to the tokenized data asset (NFT), 50\%  is  split proportionally between economic token holders and the remaining 50\% is split proportionally between ownership token holders (step 8a-8b, Fig.~\ref{fig:ecosystem}).
	The  distribution of funds are enforced by the rules in the NFT smart contract.

	\section{Discussion}
	\label{sec:discussion}

	\subsection{Economic Incentives for Token Holders}
	\label{ssec:incentives}
	
	Each data asset (NFT) is issued a corresponding pair of  ownership tokens and economic tokens (see Fig.~\ref{fig:shares3}).
	A data licensing model is used for the services provided to the consumers. Each consumer pays the NFT contract address for the licence to use the information (see step 7c, Fig.~\ref{fig:ecosystem}).
	Tokenized cash (such as stablecoin) dividends are given to economic and ownership token holders.
	For this, the payments made to the NFT contract (in step 7c, Fig.~\ref{fig:ecosystem}) are distributed as dividends (see steps 8a-8b,  Fig.~\ref{fig:ecosystem}).
	This ensures economic token holders are guaranteed payments, each time a consumer pays the NFT smart contract address --- given that each individual data asset corresponds only to the goods/services corresponding to a small portion of a building.
	Hence, the distribution is based on gross income instead of net income.
	The distribution rules are enforced by smart contracts.  
	The economic token holders, proportional to their token holdings are given a tokenized cash dividend payout at 50\% of the funds paid by the consumers.
	The remaining 50\% is distributed proportionally between the data asset owner(s), identified via ownership tokens.
	Here, owners are incentivised to hold on to their ownership tokens and work to bring in more revenue because they receive a dividend payment only when they hold the ownership tokens.
	This rule is enforced by the smart contract.

	\subsection{Decentralised Exchange (DEX)}
	\label{ssec:search}
	A DEX is a fully decentralised exchange running on a smart contract (e.g., a ERC20 DEX running on the Ethereum blockchain).
	It  is incorporated within the building ecosystem to provide trustless transfer of tokens between the buyer and seller.
	The tokens minted by the tokenizer for a data asset (see step 4, Fig.~\ref{fig:ecosystem}) is a NFT and its corresponding fungible tokens --- economic and ownership tokens.
	These tokens are transferred to the building owner(s) as part of the legal contract  with the tokenizer.
	The NFT  is used to point to the location of the information on a datastore and allows consumers to make payment transfers to the NFT contract address.
	Only the fungible tokens are sold on the DEX.
	The building owner may sell a fraction or all of her ownership stakes (steps 5a-5c, Fig.~\ref{fig:ecosystem}) by selling ownership tokens on a DEX.
	She may also raise capital by selling the economic tokens (steps 6a-6c, Fig.~\ref{fig:ecosystem}).
	The funds collected are transferred to the seller (owner) via the DEX.
	To ensure the ownership sales are only between building owners or those closely associated with the building, only whitelisted stakeholders are able to buy ownership tokens from the DEX.  
	There are no restrictions on the sale of economic tokens because it is intended for members of the public. I.e., individual and retail investors.
	
	\subsection{Search Platform}
	\label{ssec:search}
	The  search results point to building data on a datastore.
	Each data asset owner maintains her building data on a datastore\footnote{Typically, a data hosting company or  service is used.} and  points the NFT to its storage location.
	To prevent working in silos, the search is required to work across the ecosystem.
	To achieve this,  data consumers interact with a public web application that acts as a search platform for building data (step 7a, Fig.~\ref{fig:ecosystem}).
	In our architecture, any user is free to use the search platform.
	If the requested data is available, the search will return the NFT contract address and the licensing cost, for the consumer to pay (step 7b, Fig.~\ref{fig:ecosystem}).
	Once the payment is confirmed, the information is released from the datastore.
	Since a consumer pays for the information (step 7c, Fig.~\ref{fig:ecosystem}), there is a clear expectation that the requested information is precisely  what the search engine would return.
	Here, the consumer is not searching for insights but for specific building data.
	The insights are derived at a later stage by processing the data retrieved from the datastore.
	In our system architecture, the contractors are required to provide the information in the prescribed data format as part of their regular work.
	This is to simplify the collection of structured building data at its source and streamline the search.

	\subsection{Data Certifier}
	\label{ssec:certifier}
	The building information is certified by relevant  accredited experts before it is sent to the datastore.
	The public keys of all accredited experts are managed and updated on a key store smart contract (not shown in Fig.~\ref{fig:ecosystem}).
	First, the digitised information is physically verified, timestamped and digitally signed by the expert and handed over to the owner.
	Next, the owner uploads the certified data to the datastore and points the corresponding NFT to its storage location.
	Further, any consumer retrieving the information from the datastore may verify the data authenticity by reading the certifier (signer) public key from the smart contract and verifying the signature on the data.  
	
	Audit is an essential requirement for a data asset (requirement R4 in Section~\ref{ssec: DataAssets}).
	The certification carried out by data certifiers is a form of audit since it checks for impairments in the data.
	This improves accountability as it allows to detect issues with the data that needs to be resolved, and communicated  to its data owners.
	Further, the data owners may liaison with the contractors to rectify the shortcomings in  data, so it may be certified as correct.
	By marking the data as audited/unaudited on a datastore, it allows consumers to pay only from audited data based on their preference.

	\subsection{Building Data Assets Example Use Cases}
	\label{ssec:usecase}
	We present 4 types of example use cases to illustrate the applications of building data assets.
	\begin{compactenum}
		
		\item \textbf{Maintenance.}
		It is a common and repeated activity in buildings. For example, additional electric wiring and wiring improvements may require owners to hire contractors for the task.
		An electric wiring diagram saves time and effort for contractors to deduce location behind plastered walls.
		The electric wiring diagram data may be turned into a data asset.
		The same applies to plumbing diagrams. It saves time and effort for contractors to deduce
		pipes located behind walls and underground.
		Having the knowledge of its location and  layout prevents accidental digging and disruption to services.
		
		\item \textbf{Analytics and insights.}
		Building data may  be used by individuals and data analytics companies/researchers.
		For example, i.) an individual who rents a new apartment may be interested in the historic utility bills (such as power and water). She may ask the owner to provide this information but the owner has no incentive to provide it. Had the utility bill been a data asset in our ecosystem, it would be possible for the tenant to search for this information on our search platform, pay for it on the blockchain and seamlessly retrieve this information from a datastore.
		ii.) For a data analytics company, information about electricity consumption combined with information about the location (which floor) and orientation (facing which side of the building?) of the units within a building may lead to insights such as --- avoid high-floor, west-facing units to avoid high electricity bills.

		\item \textbf{Finding reusable and recyclable building materials.}
		Resource availability and allocation are an important part of resources scheduling  in  construction.
		The volume of discarded materials ending up in landfills without the intention of reuse or recycling has put strain on finding new resources afresh.
		As a  result, consumers are increasingly interested in the recyclable and reusable information at the end of the building life~\cite{Honic2021}. This along with the disassembly information~\cite{Sanchez2021}, location and availability data would give the necessary information on the  timeline for allocating the resources for new construction/renovations.
		This information may be converted into a data asset.
		
		\item \textbf{Temporary works.}
		Another potential application is in temporary works~\cite{HKICM2018part1,HKICM2018part2} (for e.g., scaffolds for access, falseworks for concrete pours, excavators to dig the earth and cranes to lift materials). It forms a large part of temporarily reused and finite resources allocated in  construction.
		While they are generally not part of the completed building, it plays an essential role in its construction.
		The knowledge of its availability is crucial in  resource allocation and avoiding delays.
		They may be tokenized into data assets in our ecosystem. Its consumers may then search for its availability, allowing us to connect the buyer and seller of the resource/service.  
		
	\end{compactenum}

	\subsection{On requirement R1. Protection from arbitrary duplication }
	\label{ssec:requrementr1}
	According to  requirement R1 (see Section~\ref{ssec: DataAssets}), it must be impossible to arbitrarily duplicate a data asset.
	R1 is an important requirement for data to be an asset. Any data that can be (easily) duplicated at will is ultimately not valuable.
	Preventing duplication of data that was viewed by a consumer is a difficult problem. Known mitigation  involves legal consequences for breaching data licensing terms.
	Hence, much of the technical focus is on preventing arbitrary duplication of  the data stored on a datastore (either via custodian or  IPFS).
	Current solutions use a combination of encryption and/or access control~\cite{venugopalan2023improving} to meet this requirement on a  datastore.
	To achieve the uniqueness requirement of the tokens used, they are issued  on a blockchain.
	The security provided by the blockchain prevents arbitrary token duplication.

	\subsection{Reasons for using a blockchain}
	\label{ssec:blockchainjustification}
	A blockchain is used for the ability to create a global ecosystem of building data assets, relevant services and businesses. It also acts as a neutral settlement layer.
	The requirements for a data asset were discussed in Section~\ref{ssec: DataAssets}.
	According to  R2, it must be possible for the owner(s) of a data asset to take possession and assume exclusive control.
	Here, strong control features are a requirement.
	The control rules embedded in the smart contract and  enforced via the consensus of the blockchain nodes are used to meet this requirement.
	According to  R3, it must be possible for a data asset to be subject to a regulatory authority (if required).
	Since we deal with data, book keeping of all transactions is necessary to meet regulatory requirements.
	Transparent and immutable book keeping is guaranteed on a blockchain.
	According to R4, it must be possible to audit a data asset (e.g., in order to verify its integrity and test for any impairments).
	Tamper-resistance and identity verification are important
	requirements for any audit.
	A blockchain is tamper-resistant under a stronger
	(byzantine adversary) threat model and all information
	is digitally signed.
	According to  R5, it must be possible to extract useful information from a digital asset, thus generating a source of revenue.
	Being able to seamlessly and securely pay and
	purchase is a necessary condition to meet this requirement.
	Trades (buy/sell) may be
	carried out on a DEX running on
	top
	of the blockchain in a trustless manner.
	While centralised solutions are faster, it may be less seamless, given its ability to block, freeze or lock assets.
	Blockchain as a  tool also meets many of the requirements of a data asset.
	For these reasons, a blockchain is used as a tool to enable a data asset.

	\section{Related Work}
	\label{sec:related}
	
	In this section, we survey existing data collection methods used in the building and construction industry.
	The work in Milojevic-Dupont et al.~\cite{Dupont2023} relied on open government datasets and a  crowdsourced volunteered geographic information database called OpenStreetMap~\cite{Arsanjani2015}.
	They built the EUBUCCO scientific database of individual building footprints for nearly 206 million buildings across the 27 European Union countries and Switzerland, together with three main attributes -- building type, height and construction year -- included for respectively 45\%, 74\%, 24\% of the buildings.
	The work in Bide et al.~\cite{Thomas2023} used  freely available data from earth observation satellites to show the Hanoi development plans far exceeded the resource planning needed to meet its completion targets.
	Other data collection techniques included the use of drones with sensors such as LiDAR~\cite{Alsayed2022} and  photographic images or other patterns of electromagnetic radiant imagery for photogrammetry~\cite{Oluibukun2021}.
	The work in Wurm et al.~\cite{Wurm2021} used deep learning based generation of building stock data from aerial images for energy demand modelling with the goal of reducing energy consumption.  
	In Heeren and Hellweg~\cite{Heeren2018}, they used two Swiss national databases to construct three-dimensional building representations.
	Enhanced GIS~\cite{Buffat2017}  including geo-spatial information (for e.g., remote sensing  satellites, drones and LiDAR) and advances in photogrammetry have increasingly assisted with building data collection.
	We observed that some of the primary sources of data are from official government datasets~\cite{HDB2023,Heeren2018}.
	Others included data from  remote sensing (e.g., flying drones~\cite{Alohan2019} with sensors
	or  satellite imagery).
	Some of the missing information was also seen to be plugged using machine learning~\cite{YANG2021} to improve the quality of data. Other machine learning techniques were used to  automatically characterise building types in urban
	areas~\cite{Ghione2022} based on publicly available image databases.
	
	\section{Conclusions and Future Work}
	\label{sec:conclusions}
	We presented our building data assets architecture to incentivise the seller (building data owner) to trade building data with the buyer (consumer of information).
	Data owners required its contractors to supply information in a prescribed digital format. Assigning data collection as part of their regular work to contractors reduced data fragmentation and a prescribed format improved data coherence.
	Consumers were able to search for the building information on a public search platform. It prevented working in silos and enabled the real-time communication of data (as and when it is updated on the datastore).
	The digitisation also removed the manual process of searching through paper copies and greatly sped up the results to the communication queries.
	Using our built ecosystem greatly improved data visibility.
	The process of building data certification acted as a form of audit and helped improve accountability.
	Building data may be conveniently searched using the platform, paid for on a blockchain and retrieved from a datastore.
	
	The work in Goy et al.~\cite{Goy2020} proposed to increase communication between data owners and data users (for e.g., through workshops and annual meetings).
	Our architecture does not consider out-of-band feedback mechanisms such as online forums or annual meetings for data owners to understand the changing requirements of consumers.  Improving communication between buyer and seller is expected to be crucial for collecting the required data and for the sustainability of the ecosystem. Out-of-band communication techniques and its incorporation are beyond the scope of the current work and is left as future work.

	\section*{Acknowledgment}
	This research is supported by the National Research Foundation, under its Campus for Research Excellence and Technological Enterprise (CREATE) Programme.
	
	\bibliographystyle{IEEEtran}
	\bibliography{ref-reduced}
	
\end{document}